\documentclass[sigconf]{acmart}%
\copyrightyear{2022}
\acmYear{2022}
\setcopyright{acmlicensed}\acmConference[Academic Mindtrek 2022]{25th International Academic Mindtrek conference}{November 16--18, 2022}{Tampere, Finland}
\acmBooktitle{25th International Academic Mindtrek conference (Academic Mindtrek 2022), November 16--18, 2022, Tampere, Finland}
\acmPrice{15.00}
\acmDOI{10.1145/3569219.3569352}
\acmISBN{978-1-4503-9955-5/22/11}
\usepackage[utf8x]{inputenc}%
\usepackage{graphbox}%
\usepackage{graphicx}%
\usepackage{makecell}%
\usepackage{multirow}%
\usepackage{tabularx}%
\usepackage{framed}%
\begin{document}%
%
\title{The Creativity of Text-to-Image Generation}%
%
\author{Jonas Oppenlaender}%
\email{joppenlu@jyu.fi}%
\orcid{0000-0002-2342-1540}%
\affiliation{%
  \institution{University of Jyv\"askyl\"a}%
  \city{Jyv\"askyl\"a}%
  \country{Finland}%
  \streetaddress{Seminaarinkatu 15}%
  \postcode{40014}%
}%
%
%
\begin{abstract}%
Text-guided synthesis of images has made a giant leap towards becoming a mainstream phenomenon. With text-to-image generation systems, anybody can create digital images and artworks. This provokes the question of whether text-to-image generation is creative.
This paper expounds on the nature of human creativity involved in text-to-image art (so-called ``AI art'') with a specific focus on the practice of \textit{prompt engineering}. The paper argues that the
current
product-centered view of creativity falls short in the context of text-to-image generation. A case exemplifying this shortcoming is provided and the importance of online communities for the creative ecosystem of text-to-image art is highlighted.
The paper provides a high-level summary of this online ecosystem drawing on Rhodes' conceptual four P model of creativity.
Challenges for evaluating the creativity of text-to-image generation and opportunities for research on text-to-image generation
in the field of Human-Computer Interaction (HCI) are discussed.
\end{abstract}%
%
%
\begin{teaserfigure}%
\centering%
  \includegraphics[width=\textwidth]{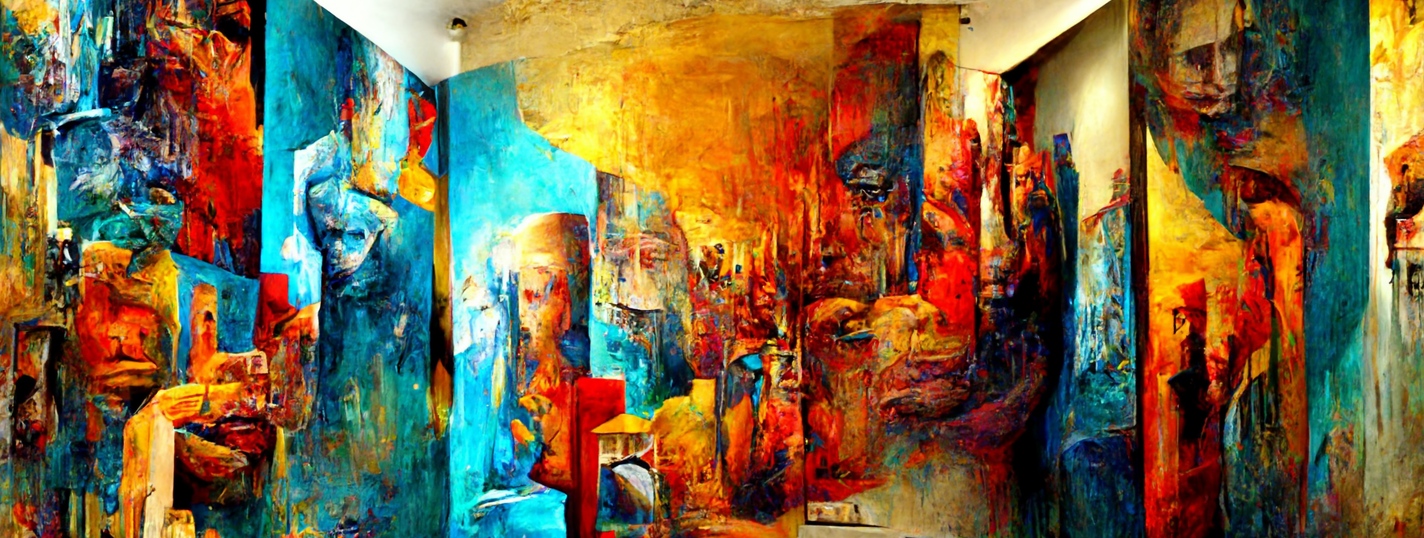}%
  \caption{Digital image generated from the prompt ``art on the wall'' with Midjourney~\cite{midjourney}.%
  }%
  \Description{A colorful digital image created from the prompt ``art on the wall'' using the text-to-image generation system Midjourney.%
  }%
\end{teaserfigure}%
%
\maketitle%
%
%
%
\section{Introduction}%
Text-to-image generation systems based on deep generative models have become popular means for creating digital images and artworks \cite{2207.13038.pdf,VQGANCLIP}.
    Given an input prompt in natural language, these generative systems are able to synthesize digital images of high aesthetic quality. 
An ecosystem of tools and resources around text-to-image generation has emerged online and a wide variety of text-to-image generation systems are now available as open source in executable notebooks.
Google's Colaboratory\footnote{https://colab.research.google.com} (Colab), in particular, is an online service that was instrumental to the growth and popularity of text-to-image generation due to Colab allowing execution of Python-based code free of charge. For this reason, Colab has become one of the main drivers of innovation in the enthusiastic text-to-image art community.
    This community consists of \textit{practitioners} (i.e., novices, amateurs, semi-professionals, but also skilled artists and professionals) with or without technical expertise.
    Many practitioners share their digital artworks on social media and in dedicated communities, such as Midjourney~\cite{midjourney}.
    Given the growing capabilities and ease-of-use of text-to-image generation systems, digital art synthesized with deep generative models is on the verge of becoming a mainstream phenomenon.

Common to text-to-image generation systems is that practitioners can create digital images and artworks with little to no understanding of the underlying technologies, simply by writing prompts in natural language (a practice called prompt engineering~\cite{gwern}, prompt programming~\cite{2102.07350.pdf}, prompt design~\cite{promptdesign}, or prompting for short).
Given the ease of use and emerging ubiquity of text-to-image generation, the question arises:
    \textbf{
    If anybody can produce digital images that resemble masterful pieces of art 
    by simply feeding textual prompts into an opaque system, is text-to-image art creative?
    What is the nature of the human creativity involved in generating images with text-to-image synthesis?
    }
In this paper, I aim to answer these questions.%

In the scholarly literature, creativity is often split into two distinct parts.
According to this two-part definition of creativity, an artifact is creative if it is both original (sometimes labeled as novel, unusual, or unique) and effective (or useful, fitting, or appropriate) \cite{2012RuncoJaegerStandardDefinition.pdf}.
This definition relates to \citeauthor{flow}'s sociocultural model of creativity in which only an idea or artifact that is deemed appropriate to the field is allowed to enter the domain~\cite{flow}. 
The two-part definition of creativity has seen vast adoption in the scholarly literature because it enables researchers to operationalize the concept of creativity into two measurable parts.
Given that measurement requires an observable product, the two-part definition is taking a product-centered view~\cite{dubina2016.pdf}. It is the outcome of the creative process (i.e., the observable artifact) that is used to determine if something is creative.

I argue that this product-centered paradigm of creativity is not enough to assess the creativity of the artifact in the case of text-to-image generation. 
This paper exemplifies a case in which the two-part operationalization of creativity fails to measure the full extent of human creativity.
Instead of only looking at the end product of the creative process, we need to expand our view to include the whole creative process.
Rhodes' conceptual model of the ``4~P'' of creativity~\cite{4P} provides a fitting framework for this investigation. The framework 
looks at creativity from four perspectives~-- product, person, process,  and press (i.e., the environment).
    In the context of text-to-image generation, all four perspectives are necessary to capture the full extent of human creativity (with special emphasis on the latter three).
    Online communities, in particular, form an increasingly important social constituent of the online creative ecosystem around text-to-image generation.

This paper expounds on the nature of human creativity prevalent in the sub-culture of text-to-image art (sometimes referred to as ``AI art'' or ``generative art'' \cite{boden2009.pdf,galanter_generative.pdf} within the online community)).
When creating art from text, practitioners cede control (in part) to artificial intelligence~(AI)~\cite{galanter_generative.pdf}. Consequently, the human creativity in text-to-image synthesis lies not in the end product (i.e., the digital image), but arises from the interaction of humans with the AI and the resulting practices that evolve from this interaction (e.g., ``prompt engineering'').
For instance, image-level and portfolio-level curation are important practices relating to the process of text-to-image generation.
These novel creative practices are shaped and informed by a growing ecosystem of community-driven resources and tools. In particular, online communities and online resources are factors with growing importance influencing the human creativity involved in text-to-image generation. 

This paper makes the following contributions:%
\begin{itemize}%
    \item A case 
    in which the popular product-based operationalization of creativity fails to measure the human creativity involved in text-to-image generation is exemplified.
    \item Rhodes' four P framework~\cite{4P} is used to expound on the nature of human creativity involved in text-to-image generation.
    Special focus is placed on the iterative and interactive practice of ``prompt engineering'' and the online community of practitioners of this novel creative practice.
    \item Image-level and portfolio-level curation are highlighted as two important creative practices involved in the creative process of text-to-image generation.
\end{itemize}%
Additionally, attention is pointed to the growing importance of communities in the emerging ecosystem of text-to-image generation as a catalyst for creativity and learning. Online communities and resources are important factors affecting the human creativity of text-to-image art.
An outline of five different roles taken by members in the AI art community is provided.
Finally, the practical challenges of evaluating the creativity of images synthesized with text-to-image generation systems are discussed and an outline of opportunities for future research is provided.

The paper is structured as follows.
After first discussing related work (Section \ref{sec:relatedwork}) and the methodology (Section \ref{sec:selfdisclosure}),
I discuss and illustrate a case in which a person creates a digital artwork in a non-creative way (Section \ref{sec:edgecase}).
Motivated by this case, I argue that the human creativity of text-to-image synthesis lies in the interaction of practitioners with the deep learning based system and the creative practices arising from this interaction, such as prompt engineering and curation.
I~expand on the nature of human creativity involved in text-to-image generation in Section~\ref{sec:creativity}.
I~then discuss opportunities for future research on text-to-image generation in the field of Human-Computer Interaction (HCI) and the broader implications of text-based co-creation with AI-based systems in Section~\ref{sec:discussion}.
The paper concludes in Section~\ref{sec:conclusion}.

\section{Related Work}%
\label{sec:relatedwork}%
\subsection{Text-to-Image Generation}%
\label{sec:TTIS}%
The text-guided synthesis of images using deep learning has made significant advances since the inception of Generative Adversarial Networks (GANs)~\cite{GAN} in 2014 and Google's Deep Dream~\cite{deepdream} in 2015.
It was the release of OpenAI's CLIP \cite{CLIP} in January 2021 that spurred immense technical progress in text-to-image generation.
CLIP is a contrastive language-vision model conceived for the task of classifying images by providing the names of the visual categories to be recognized.
CLIP was trained 
on a large corpus of images and text from the World Wide Web.
Due to the Web-scale size of its training data, the CLIP model has
learned a wide variety of visual concepts from natural language supervision and can associate them with their names.
Given an image and a set of labels (e.g., ``a photo of a dog''), CLIP can predict the label that is most likely to be paired with the image.
It is this ability to visually associate natural language with images that resulted in CLIP being used in generative systems.

In the context of text-to-image generation, CLIP found its first significant application in GAN-based image generation systems.
When used as a discriminator component in a generative deep learning architecture,  CLIP can guide the generator component to produce digital images that best match a given textual prompt.
Shortly after OpenAI released CLIP in January 2021, AI enthusiasts created GAN+{\allowbreak}CLIP based systems for the specific purpose of generating digital art.
Ryan Murdoch
created ``Big Sleep'' \cite{bigsleepgithub,bigsleepiccc}, a combination of a GAN called BigGAN and CLIP.
This inspired Katherine Crowson
to connect an even more powerful neural network (VQGAN) with CLIP~\cite{VQGANCLIP}.
The combination of VQGAN and CLIP was very popular in 2021 and became one of the standard techniques of generating artworks until the method was superseded by diffusion-based systems~\cite{NEURIPS2021_49ad23d1}.
One can argue that VQGAN--CLIP was instrumental to advancing the emerging field of text-to-image art~\cite{VQGANCLIP}. The source code of VQGAN--CLIP was available online, and many generative systems and methods have since been developed based on the pioneering work of Murdoch and Crowson.
Today, practitioners can choose from a growing variety of systems for generating text-to-image art. The current state of the art are diffusion models, such as CLIP guided diffusion~\cite{CLIPguideddiffusion} and latent diffusion~\cite{latent-diffusion}.

A growing number of generative systems are available as open source in Jupyter notebooks on Google's Colaboratory\footnote{https://colab.research.google.com} (Colab), on Huggingface\footnote{https://huggingface.co}, and other places online where they can be downloaded or executed free of charge and with relatively high ease of use.
Due to this low barrier of entry, Colab notebooks contribute to a democratization of digital art production~-- anybody can create digital images and artworks with text-to-image generation systems.
This raises the question about the level and nature of human creativity involved in text-based generative art (or ``AI art'' as it is often called within the community of practitioners).
This paper draws on Rhodes' 
model of creativity as an analytical framework to answer this question, as outlined in the following section.%
%
%
%
\subsection{The Four P of Creativity}%
\label{sec:4p}%
%
Creativity is a complex multi-faceted phenomenon that is difficult to define \cite{PHD}.
In the field of Human-Computer Interaction (HCI), a multitude of definitions of creativity are in use which has contributed to a ``fuzziness'' around the concept of creativity in the scholarly literature~\cite{p1235-frich.pdf}.

One important model of creativity is \citeauthor{flow}'s ``systems model'' of creativity~\cite{flow}.
The model consists of three parts: the individual person having an idea or creating an artifact, the domain (embedded in the broader context of culture), and the field which consists of individuals (gatekeepers) assessing the creativity of the idea or artifact before it enters the domain.
According to this sociocultural model of creativity, ``social confirmation is necessary for something to be called creative'' because ``creativity does not happen inside people’s heads, but in the interaction between a person’s thoughts and a sociocultural context'' \cite{flow}.
Contrary to the above, one can argue that creativity can also take place on a personal level when an individual creates something personally valuable and meaningful (so-called small-c creativity \cite{Beyond_Big_and_Little_The_Four_C_Model_of_Creativi.pdf} or everyday creativity \cite{Richards}).
This is opposed to Big-C (or eminent) creativity which requires a large contribution to society for something to be considered creative (e.g., a Nobel-prize winning discovery) \cite{Beyond_Big_and_Little_The_Four_C_Model_of_Creativi.pdf,Richards}.

The above model and definition of creativity are coarse and do not capture the multitude of activities that lead to the creation of a creative artifact.
One useful conceptual model for reflecting on the entirety of the concept of creativity is Rhodes' model of the ``four~P'' of creativity~\cite{4P}.
The model aims to explain what it means to be creative by looking at four components.
\textit{Person} pertains to a human beings' personality, intellect, habits, attitudes, and other factors that affect a person's creativity. 
\textit{Process} pertains to the process of creating ideas and artifacts (e.g., thinking, motivation, learning, etc.).
\textit{Press} refers to the human being's environment which can influence the person and the mental processes.
Last, \textit{product} pertains to ``artifacts of thought'' -- ideas that have been communicated to other people in observable form (e.g., as a poem, painting, or sculpture)~\cite{4P}.

As in the above sociocultural model and the two-part definition of creativity, all four aspects in Rhodes' conceptual model are linked to an observable product~\cite{dubina2016.pdf}.
    The underlying assumption is that the embodiment of an idea~-- i.e., the outcome of a process, such as a painting, poem, sculpture, or an idea scribbled on a napkin~-- is a proxy for the creativity of a person.
    Following this perspective, the observable product can serve as a measure for human creativity.
One popular operationalization of this measure in the scholarly literature is the two-part ``standard'' definition of creativity \cite{2012RuncoJaegerStandardDefinition.pdf,plucker2004.pdf}. Following this definition, an artifact needs to be both novel (original, unique, etc.) and effective (appropriate, valuable, useful, etc.) to be considered creative.
According to this definition, it is therefore not enough for a product to be novel, it also needs to be appropriate, useful, or valuable for someone.
    Following the Big-C perspective of creativity, this someone is a larger group of people or society as a whole (the gatekeepers in the field), and
    from a small-c perspective, it is the creator of the creative artifact~\cite{Beyond_Big_and_Little_The_Four_C_Model_of_Creativi.pdf}.%
%
%
\section{Method}%
\label{sec:method}%
\label{sec:selfdisclosure}%

As 
self-disclosure to contextualize this paper, I first provide details about my research on text-to-image generation and my background
before I describe the research method.

\subsection{Researcher's background}%
I am a Computer Scientist with a background in Human-Computer Interaction (HCI) and Human-Centered Computing (HCC).
My past research interest lies in \textbf{human creativity}, not computational creativity \cite{ComputationalCreativity.pdf}.
What interests me in text-to-image generation is the interaction of humans with artificial intelligence (AI) and the lived and internalized practices that may arise from repeated interactions with AI.
While I have created several thousands of images with text-to-image generation systems (as explained in the remainder of this section), I do not consider myself an artist and I pursue text-to-image generation merely for intrinsic pleasure.%

\subsection{Online Ethnography}%
This work is grounded in an online ethnography \cite{Netnography} of the text-to-image art community conducted between October 2021 and March 2022.
I discovered and learned about text-to-image art in mid-2021 in the postings from Katherine Crowson (@Rivers{\allowbreak}Have{\allowbreak}Wings) and others on Twitter.
Since October 2021, I have been experimenting with text-to-image generation systems. I initially started to experiment with VQGAN--CLIP due to
    its performance, ease of use, and popularity (at the time).
    Later, I also experimented with CLIP-guided diffusion and other systems available on Google Colab.
For learning about writing effective prompts, I turned to Twitter. However, there is a natural limit to the learning curve on Twitter: since the full prompt is often not shared by practitioners on Twitter, one is left none-the-wiser how the digital artworks were created.
The launch of the Midjourney\footnote{https://www.midjourney.com} community proved to be a eureka moment in this regard.
Members of the Midjourney community generate images by typing prompts into Discourse-based chat rooms. The images are displayed shortly after in the chat. Members can see the images created by other members together with the respective prompts.
Therefore, the prompts are shared in the community which makes the community a vast social learning resource, unlocking the social creativity of what previously was a lone-creator practice.
Communities like Midjourney are a game-changer in advancing the practice of prompt engineering for text-to-image generation.
I was invited into the Midjourney beta program in mid-March 2022, 
and have since practiced and studied how prompt engineering for text-to-image art is conducted in practice.

Drawing on Rhodes' multiperspective model of creativity, my research seeks to develop a broad overview of the emerging creative ecosystem of text-to-image synthesis and text-based generative art (AI art) to find answers to the research questions mentioned in the introduction. This ecosystem consists of online resources and online communities with many participants. There are thus many factors affecting the creativity of practitioners in these communities. Given this complexity, this work takes a pluralistic research approach, combining different qualitative methods.
Following a participant-observational approach \cite{Netnography},
online communities (Twitter and Midjourney) were studied.
Part of this work is based on the prompt writing practices that I observed, first on Twitter and later in the Midjourney community.
Additionally, resources emerging from the online community (e.g., guides and other helpful resources created by practitioners) were studied and reviewed.
The review also involved both scholarly literature, particular on prompt engineering and text-to-image generation, and gray literature (e.g., blog posts and news articles).
Data was gathered in field notes which were iteratively revisited and extended.

\section{Feeding Random Snippets of Text to the Machine}%
\label{sec:edgecase}%
%
\newlength{\bigimgwidth}%
\setlength{\bigimgwidth}{.4\textwidth}%
\newcolumntype{Y}{>{\centering\arraybackslash}X}
\begin{figure*}[!ht]
\centering
\begin{tabularx}{\textwidth}{YY}%
    \includegraphics[width=\bigimgwidth]{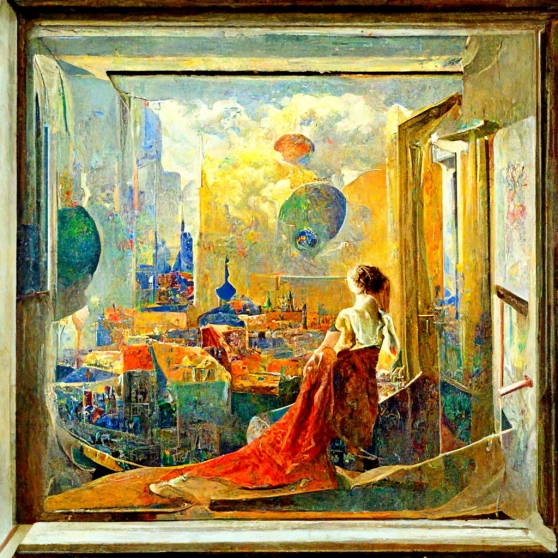}%
 &%
    \includegraphics[width=\bigimgwidth]{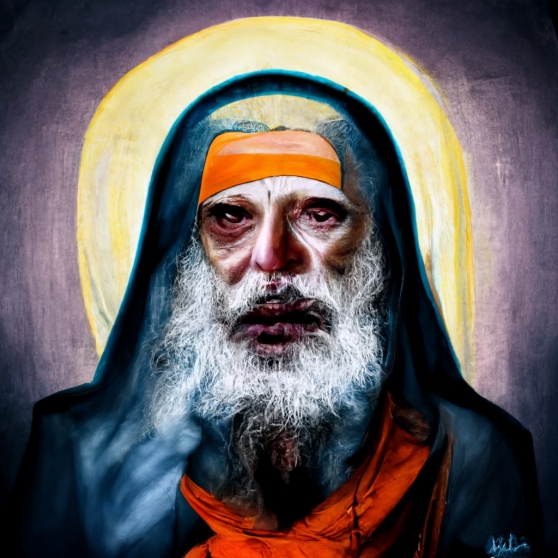}%
\\%
    \textit{What if everything around you isn't quite as it seems What if all the world you used to know Is an elaborate dream. Painting by Haralampi Oroschakoff, trending on wikiart}\newline%
    (lyrics from the song `Right Where It Belongs' by Nine Inch Nails)%
&
    \textit{How long will this take, baba? And how long have we been sleeping? Do you see me hanging on to every word you say? How soon will I be holy? How much will this cost, guru? How much longer till you completely absolve me?}\newline%
    (lyrics from the song `Baba' by Alanis Morissette)%
\vspace{1em}
\\
    \includegraphics[width=\bigimgwidth]{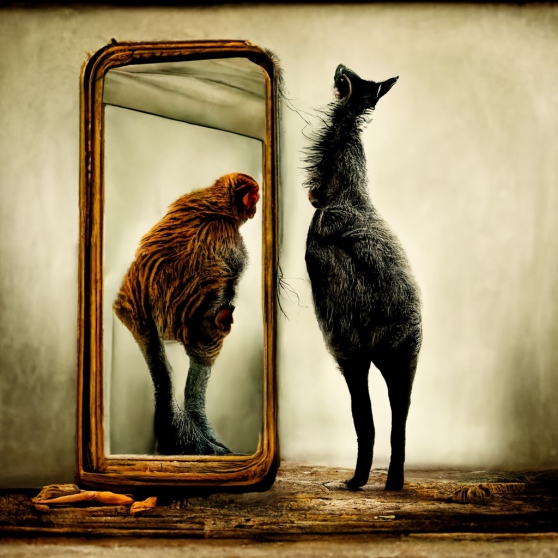}
 &
    \includegraphics[width=\bigimgwidth]{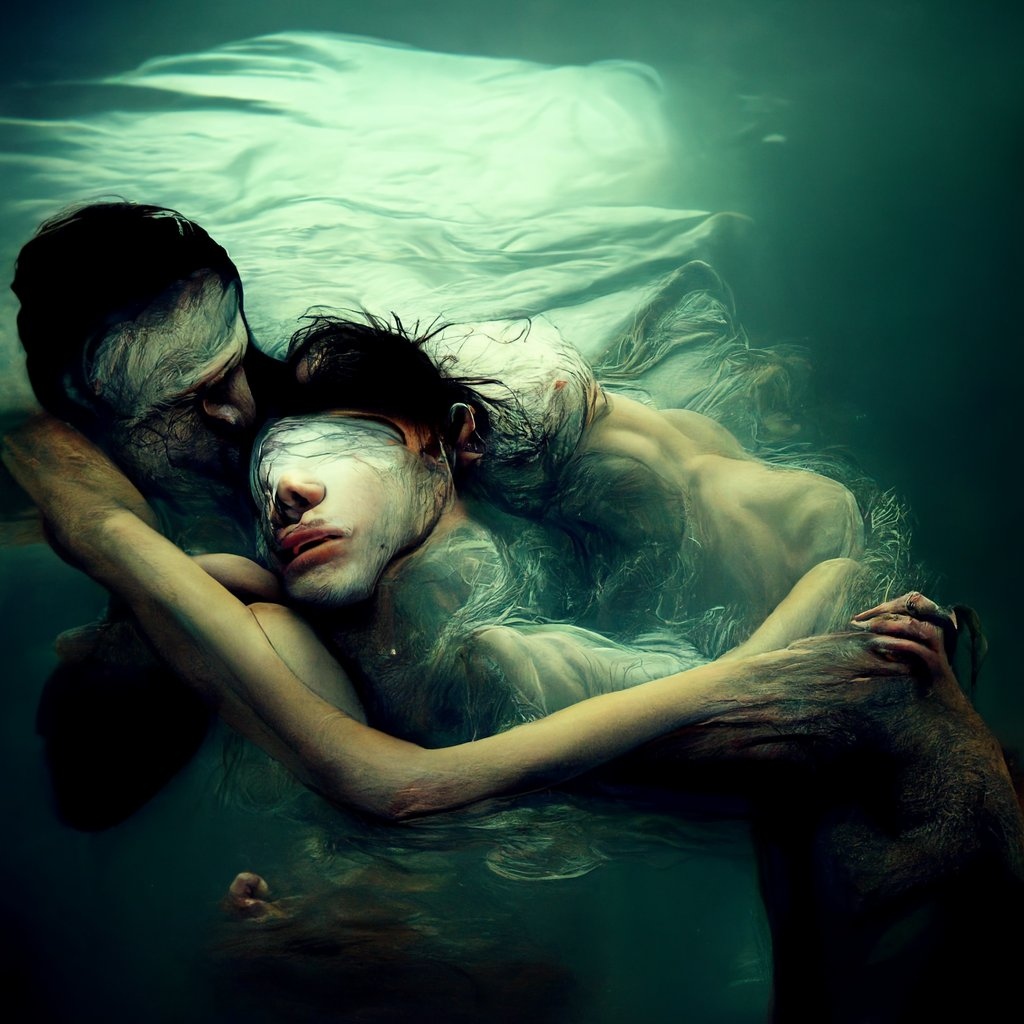}
\\
        \textit{When I see the man in the mirror I see an animal clearer}\newline%
        (lyrics from the song `Hot' by CunninLynguists)
&
        \textit{Soaked in soul, he swims in my eyes by the bed}\newline%
        (lyrics from the song `Wake Up Alone' by Amy Winehouse)
\end{tabularx}%
  \caption{Examples of images generated from parts of music lyrics with Midjourney's CLIP guided diffusion \cite{midjourney}.
  Figure a) demonstrates the use of additional prompt modifiers.
  Figures b) and c) were created without prompt modifiers.}%
  \Description{Examples of images generated from parts of music lyrics.}%
  \label{fig:examples1}%
\end{figure*}%
In the context of text-to-image synthesis, the product-centered definition of creativity is not enough to fully assess the human creativity involved in creating a digital artwork.
The whole ``four P'' of creativity are needed to capture the extent of the human and social creativity involved in text-to-image generation.
In this section, I motivate and illustrate this point with a case in which the product-centered definition of creativity fails to capture the full extent of the creativity involved in text-to-image generation.



Text-to-image generation systems have become exceedingly good at potentially turning any given textual input into high-fidelity images, no matter what input is given. For instance, consider the following two scenarios:%
\begin{framed}%
\begin{quote}%
    (1) A person takes a random snippet of text, such as words from an encyclopedia, and feeds the words to a text-to-image generation system.
    \vspace{.5\baselineskip}
    \\
    (2) A person listens to music and types verbatim lyrics into a text-to-image generation system.%
\end{quote}%
\end{framed}%
In the latter case, the information passes through the person's perceptual system where it is, of course, subject to interpretation, misinterpretation, bias, and other conscious and unconscious processes.
But assuming that the person correctly understands the music lyrics and merely transmits them to the text-to-image generation system, \textit{is this creative?}


\newlength{\imagewidth}%
\setlength{\imagewidth}{.18\textwidth}%
\newcommand{\mytd}[1]{\small{\multicolumn{1}{c}{#1}}}
\newcolumntype{Y}{>{\centering\arraybackslash}X}
\begin{figure*}[!ht]
\centering
\begin{tabularx}{\textwidth}{YYYYY}
    \multicolumn{2}{c}{\textbf{Single characters}}
       &
    \textbf{Single words}
       &
    \textbf{Movie quotes}
       &
    \textbf{Emojis}
\vspace{.5em}
\\
    \includegraphics[width=\imagewidth]{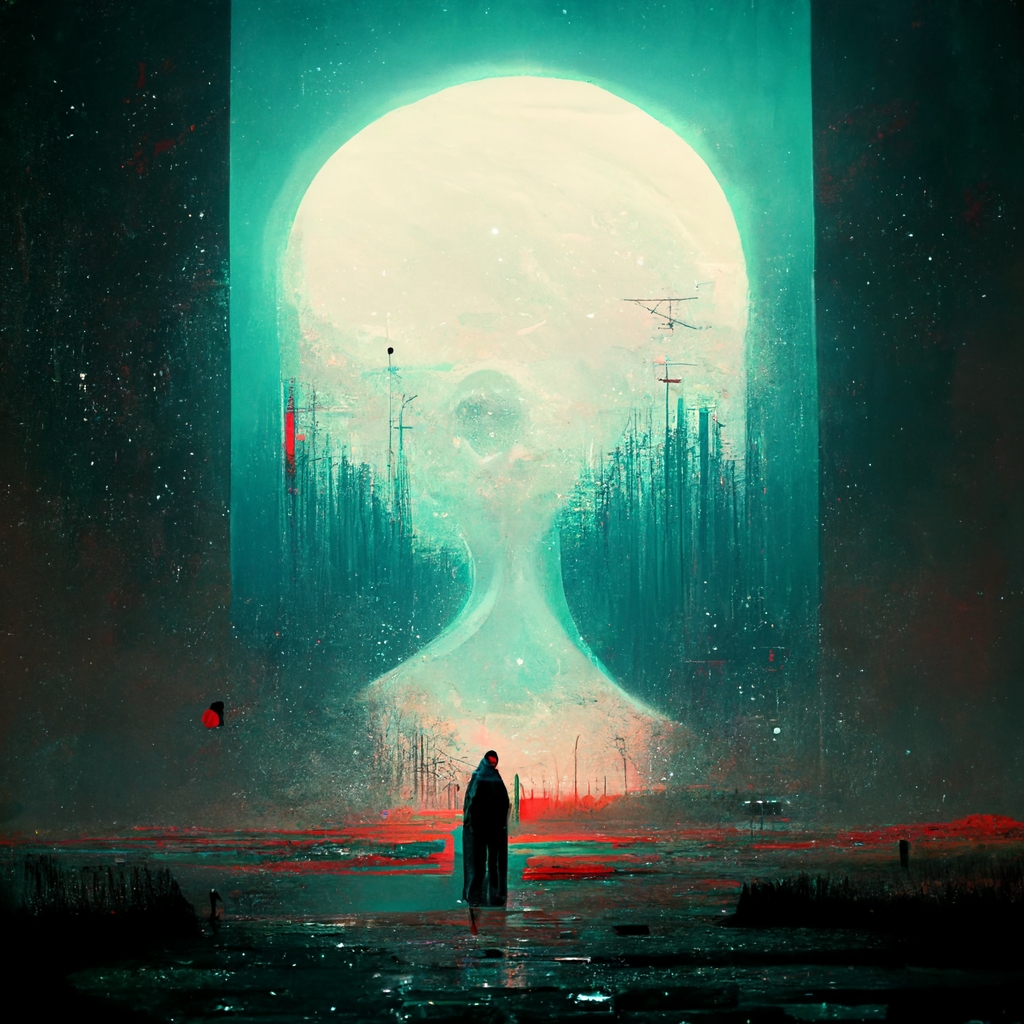}
       &
    \includegraphics[width=\imagewidth]{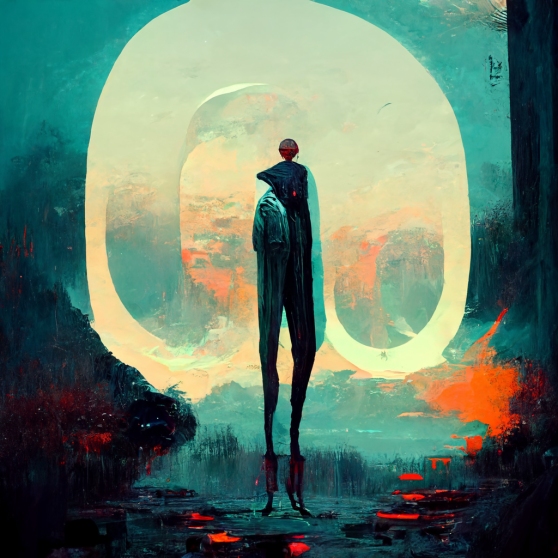}
       &
    \includegraphics[width=\imagewidth]{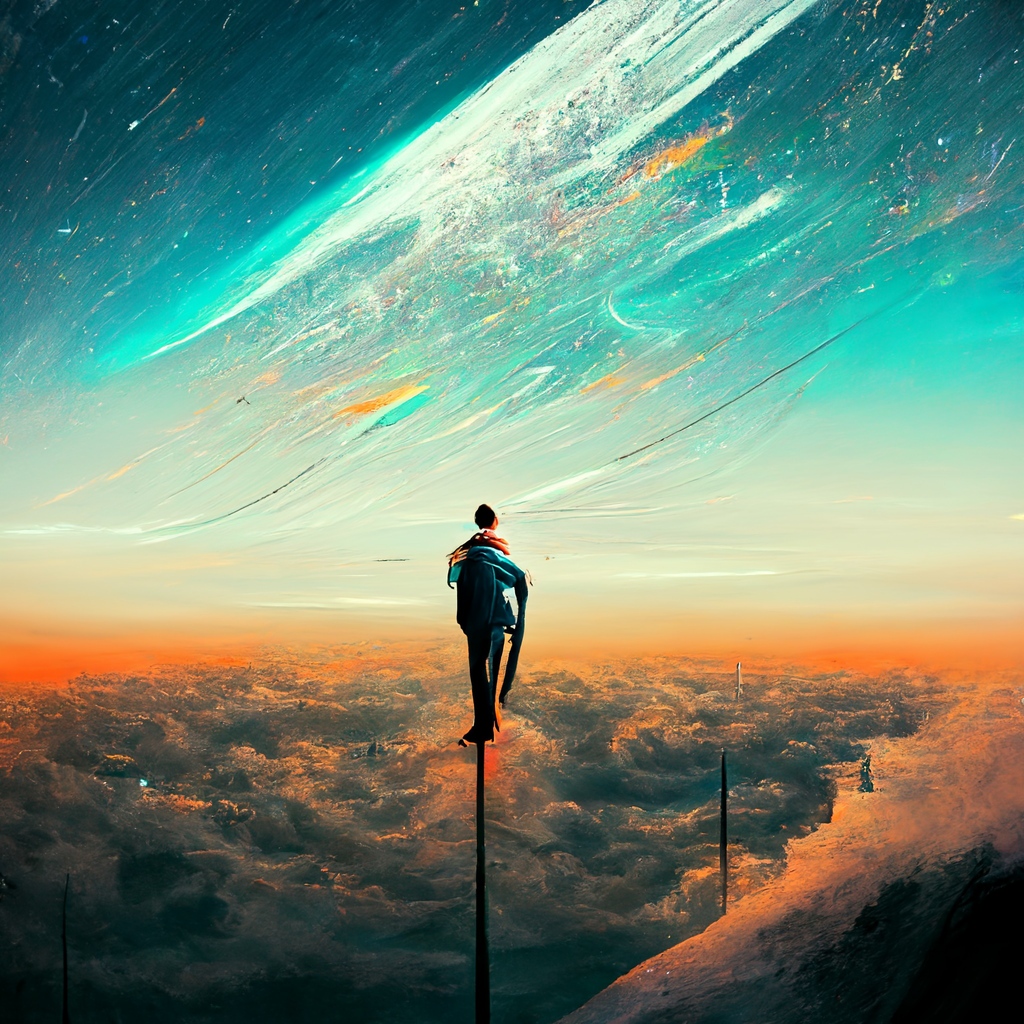}
       &
    \includegraphics[width=\imagewidth]{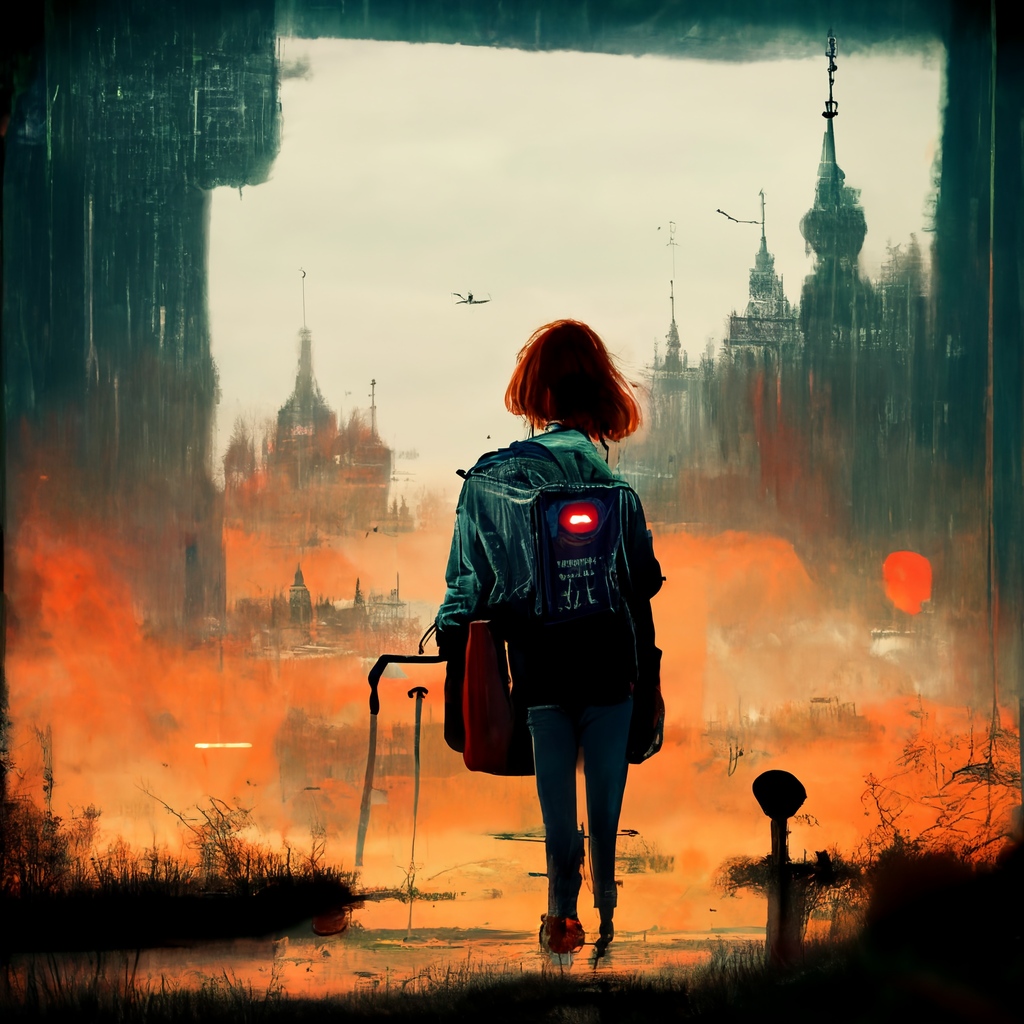}
       &
    \includegraphics[width=\imagewidth]{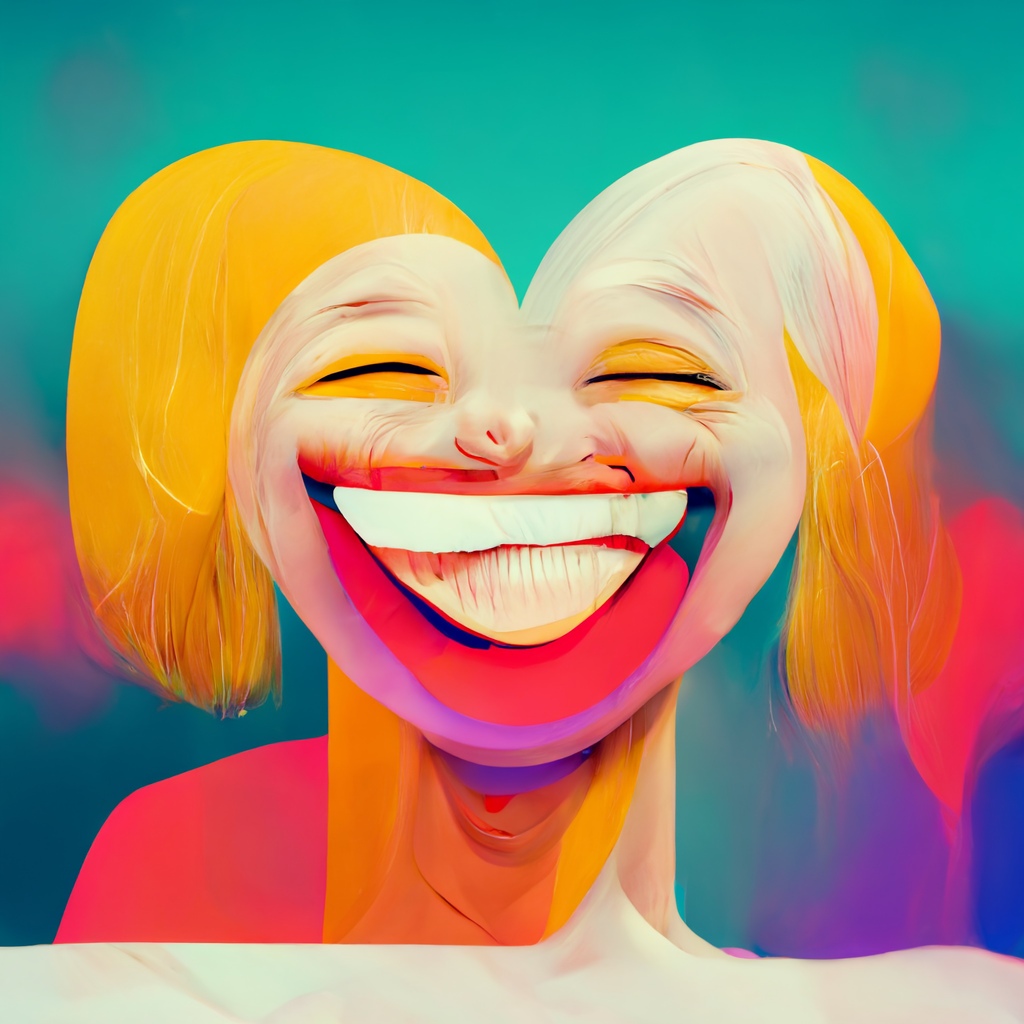}
    \\
        a) $\Delta$
       &
        b) O
       &
        c) \textit{limitless}
       &
        d) \textit{I'll be back}
       &
        e) 
        \includegraphics[width=11pt,height=11pt,align=c]{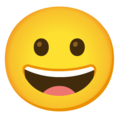}
\vspace{.5em}
\\
      \includegraphics[width=\imagewidth]{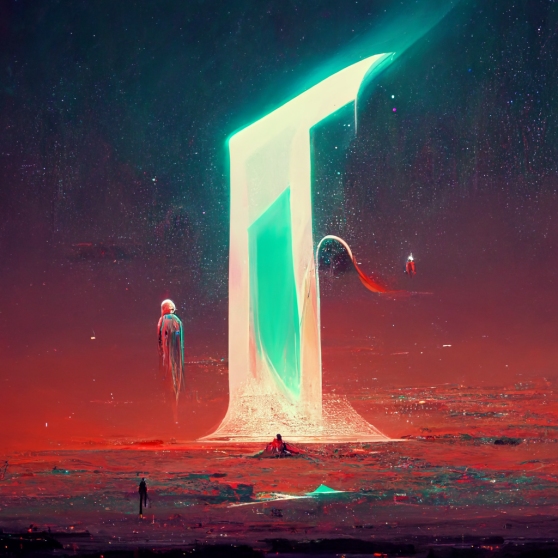}
       &
    \includegraphics[width=\imagewidth]{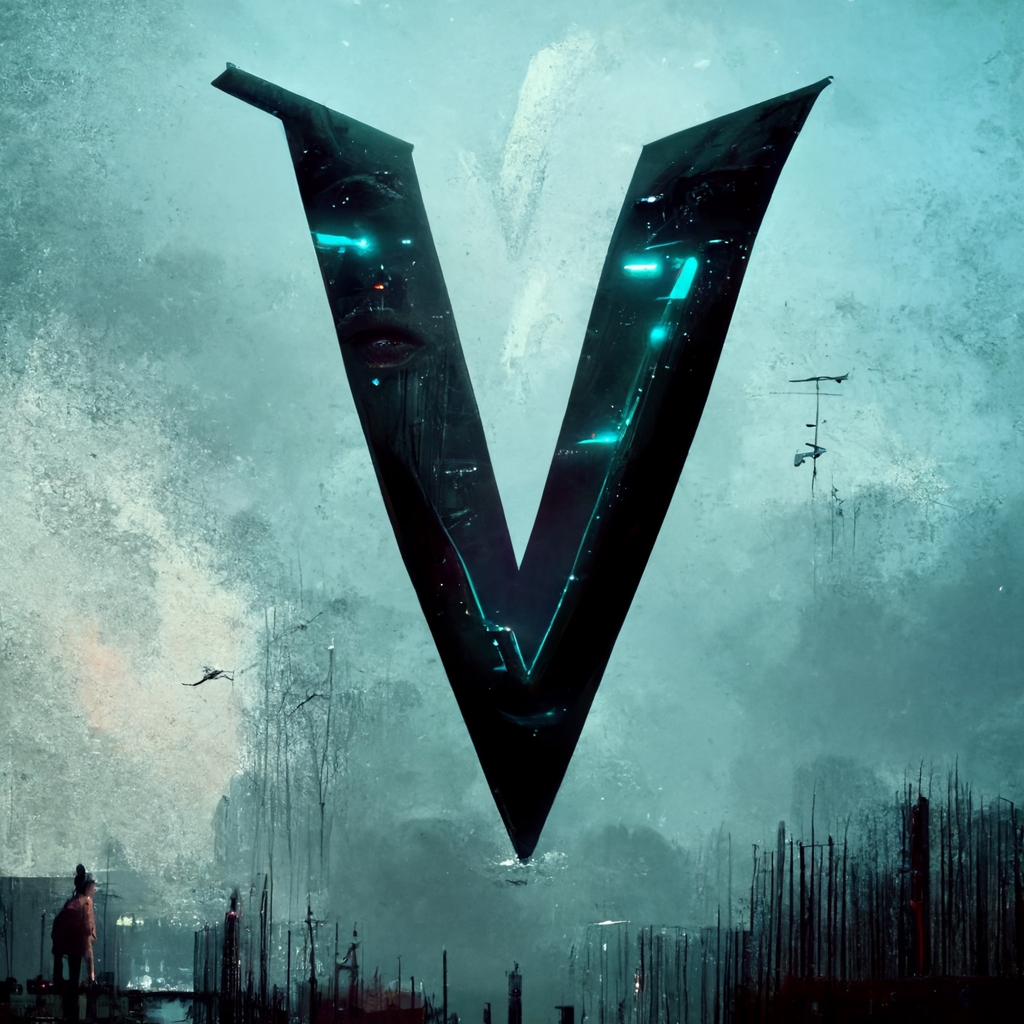}
       &
    \includegraphics[width=\imagewidth]{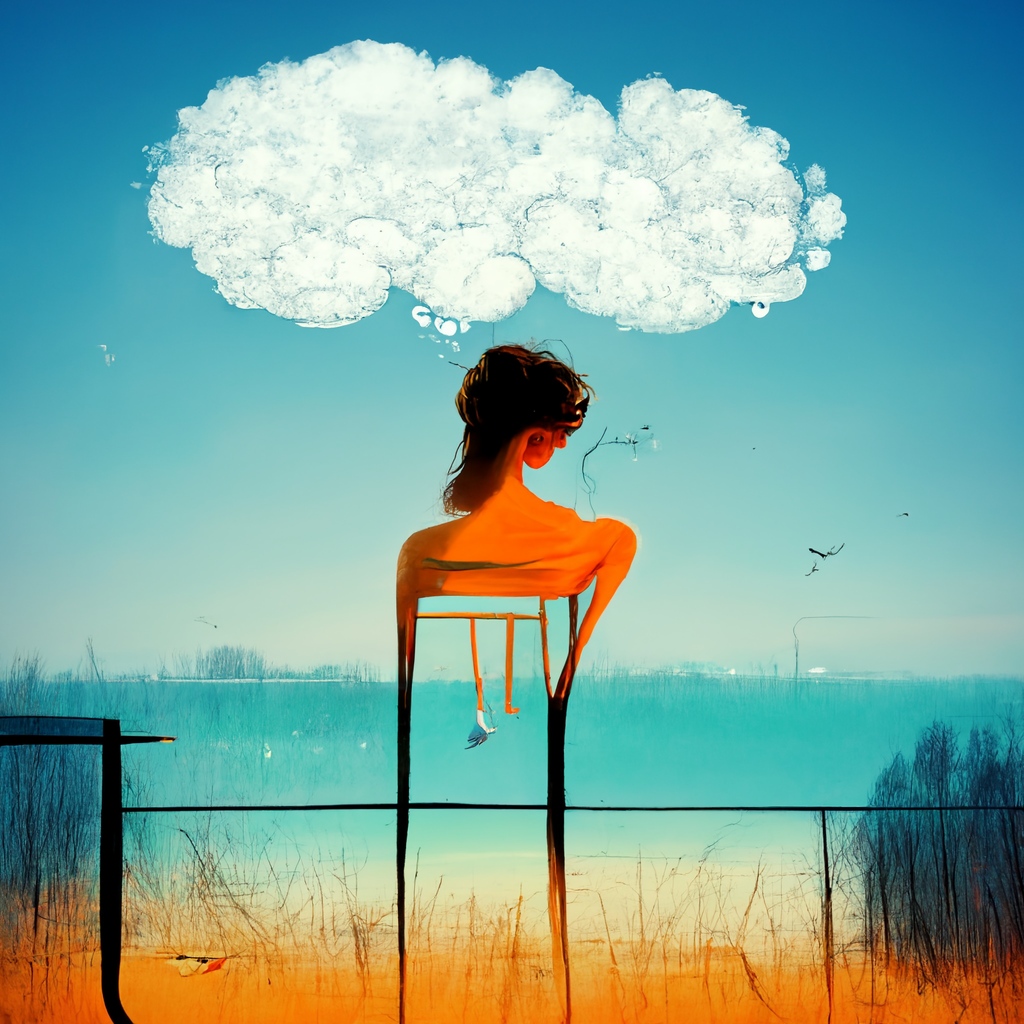}
       &
    \includegraphics[width=\imagewidth]{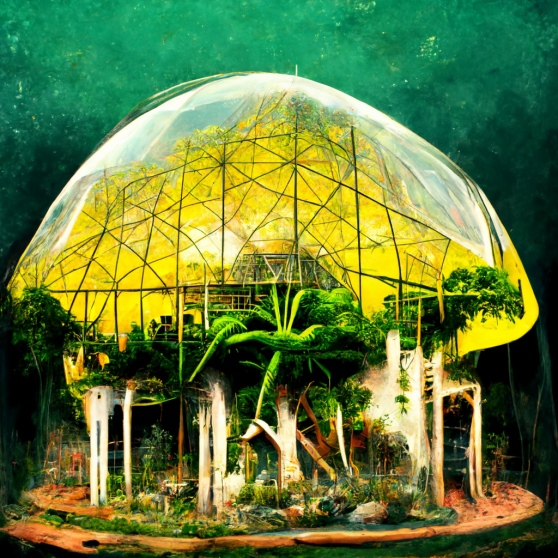}
       &
    \includegraphics[width=\imagewidth]{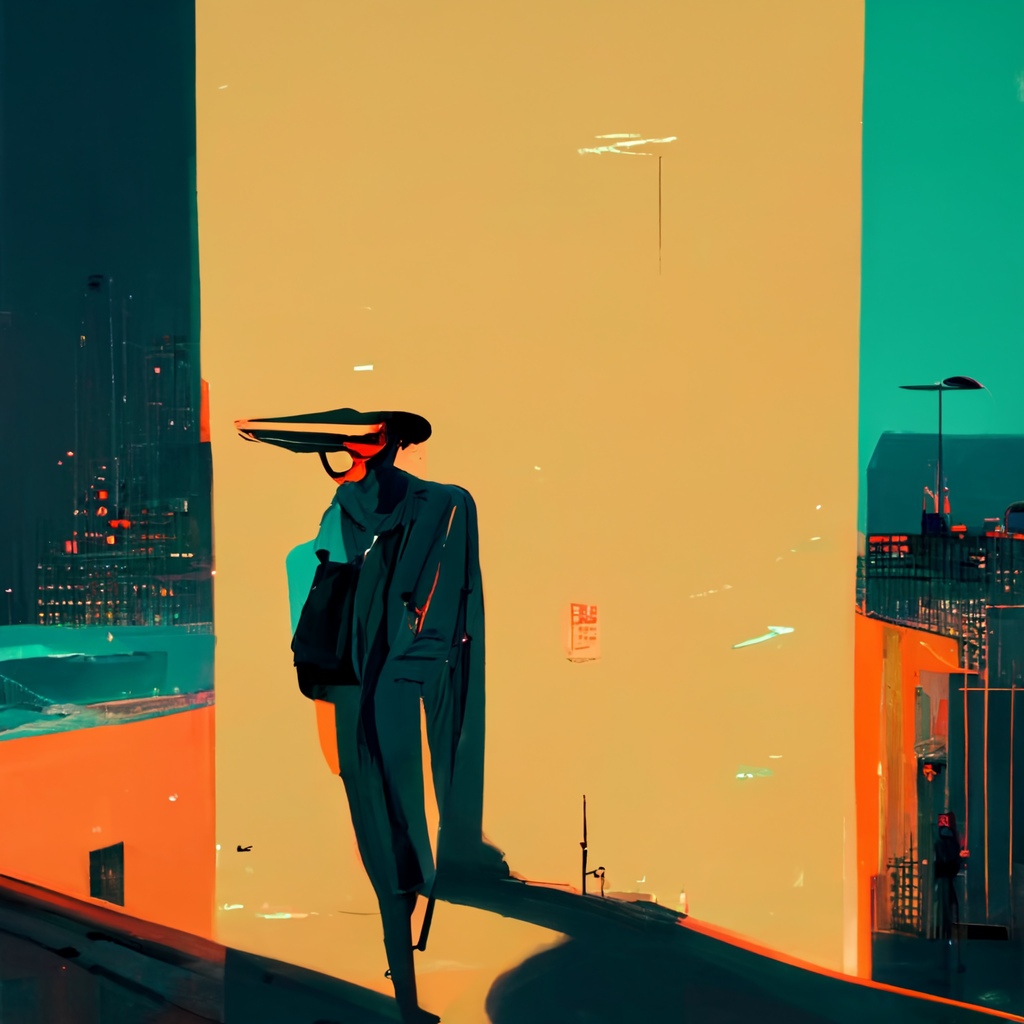}
    \\
        f) $\infty$
       &
        g)  V
       & 
        h) \textit{daydreaming}
       & 
        i) \textit{viva los} [sic]
           \newline
           \textit{bio dome}
       & 
        j)  
        \includegraphics[width=11pt,height=11pt,align=c]{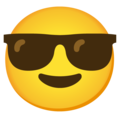}%
\\%
\end{tabularx}%
  \caption{Examples of images generated with Midjourney CLIP guided diffusion without style modifiers~\cite{midjourney}.}%
  \Description{Examples of images generated with Midjourney CLIP guided diffusion without style modifiers.}%
  \label{fig:examples2}%
\end{figure*}%
%

These two scenarios serve to illustrate that little to no human creativity 
may be involved in producing art with text-to-image synthesis.
The second scenario is comparable to Searle's thought experiment of the ``Chinese Room''~\cite{searle_1980}. In this philosophical argument, a person receives a task while being locked in a room with no access to information from the outside world and, by following detailed instructions, produces sophisticated outputs without having any specific skills or knowledge.
Similarly in the case of text-to-image generation, little or no creativity may be involved besides some basic literacy skills.

I have extensively experimented with entering music lyrics into text-to-image generation systems ({VQGAN--CLIP}, CLIP Guided Diffusion, and Midjourney). From this personal experience and observations in the Midjourney community, I can tell that surprising, interesting, and high-quality artworks can be produced from lyrics and random textual inputs.
    Figures \ref{fig:examples1} and \ref{fig:examples2} illustrate this point with hand-picked examples of textual input prompts and their respective images. The images in Figure \ref{fig:examples1} were created from music lyrics. The images in Figure \ref{fig:examples2} were created with single characters (figures a, b, f, and g), single words (c and h), short quotes from movies (d and i) and emojis (e and j).

The figures exemplify that practitioners do not necessarily need to stretch their imagination and exercise their creativity to produce high-quality results  
with state-of-the-art text-to-image generation systems. In this case, the interaction with the system is both skill-free (besides some basic literacy) and not creative.
The end product~-- i.e., the generated digital image~-- is an imperfect proxy for the person's creativity.
Much of the recent criticism of text-to-image synthesis voiced by artists (e.g., \cite{artisdead}) can be predicated on this point. Text-to-image synthesis allows hobbyists to create digital artworks in seconds as compared to dozens or even hundreds of hours using traditional methods.
The case presented in this section serves as motivation to expand upon the human creativity involved in text-to-image generation in the following section.%
%
%
\section{The Human Creativity of Text-to-Image Generation}%
\label{sec:creativity}%
%
I argue that the creativity of text-to-image generation arises from  the text-based interaction of human users with text-to-image generation systems and the human-computer co-creativity that is determined by the constellation of user and AI~\cite{062-iccc20.pdf,3519026.pdf}.
Increasingly, online communities are also becoming determining factors in how users find inspiration and learn the craft of prompt engineering. It is this increasingly social practice of prompt engineering 
that makes text-to-image generation creative.
In the following, Rhodes' 4~P model serves as a framework to expand upon the human creativity involved in text-to-image generation.


\subsection{Product: The Digital Image}
Text-to-image generation with the aim of synthesizing digital art has made significant advances in 2021. The state-of-the-art of text-to-image generation systems are able to synthesize images of high aesthetic quality, especially if prompt modifiers are being used to alter the style or boost the quality of the image (c.f. Figure \ref{fig:examples1} and~\cite{chilton,modifiers}).
From the product-based perspective of creativity, the resulting images are, of course, creative in themselves.
But, as exemplified in Section \ref{sec:edgecase}, the images may not be the fruit of human creativity. Instead, the images are a result of computational creativity~\cite{ComputationalCreativity.pdf}.

Today's artworks created with text-to-image synthesis are still not perfect, especially if human anatomy and faces are depicted.
    Additional post-processing steps are often necessary in these cases (e.g., to correct the eyes of human subjects).
But it can be expected that we will, in the near future, no longer be able to tell text-to-image artworks apart from creations made by humans \cite{passed-uncanny-valley}. Therefore, text-to-image artworks (i.e., the ``product'' of text-to-image generation systems) are an imperfect measure for human creativity. Instead of looking at the image as creative artifact, we need to turn to the other three~P in Rhodes' model (person, process, and press) to assess the extent of human creativity involved in text-to-image generation.%
%
%
\subsection{Person: The Practitioner}%
\label{sec:person}%
In my study of and participation in the Midjourney community, I observed some practitioners produce an abundance of creative, innovative, and exotic prompts,
while others seem to struggle or feel inhibited.
    For instance, it can be observed that some novice practitioners
    attempt (and often fail) to coax images out of the text-to-image generation system with very long and specific prompts, while others effortlessly produce beautiful images with rather minimalistic prompts.

Writing effective prompts is a skill linked to a person's 
knowledge of the training set and the neural networks' latent space, but also the person's knowledge of and experience with
different prompt modifiers \cite{modifiers}.
Together, this knowledge and the skills constitute the practice of
``prompt engineering''~\cite{gwern,chilton}~-- that is, the creative practice of writing effective textual input prompts for text-to-image generation systems.
    This skill also goes beyond mere knowledge of textual prompts. For instance, knowledge of which aspect ratio to choose for a specific subject and an understanding of the system's training data and configuration parameters is key to produce high-fidelity images.
Prompt engineering is a learned skill, because it is not immediately apparent how to write effective prompts and which keywords make good prompt modifiers \cite{modifiers}.
    For instance, one popular prompt modifier used in the text-to-image art community is ``trending on artstation.'' The community of practitioners found that adding this modifier to a prompt will increase the quality of generated images. Many other prompt modifiers that modify the style and quality of generated images have since been discovered and are regular being applied in the community.
The following section highlights the iterative nature of applying prompt modifiers to prompts.

\subsection{Process: Iterative Prompt Engineering and Image Curation}%
\label{sec:process}%
The creative process is vast and may involve a multitude of different workflows, practices, and tools.
The current view of image generation systems, as, for instance, held by Emad Mostaque (the founder of Stability.ai) is that text-to-image generation models are only the first part in a creative pipeline that extends over multiple AI-based systems and applications \cite{pipeline}.
A practitioner may, for instance, generate a first sketch with Stable Diffusion \cite{latent-diffusion}, run the output through a model that was trained on improving faces (such as GFPGAN \cite{wang2021gfpgan}), and then proceed to perform more advanced techniques (e.g., inpainting and outpainting) before finally editing the result in a graphics editor.
In this section, I want to focus on two parts of the creative process that I deem to be native to text-to-image generation:
iteration and curation.

\subsubsection{Iterative prompt engineering}%
Text-to-image art is driven by exploration.
Practitioners probe the model's latent space with textual prompts to see what works (and what not)~\cite{aliendream}.
This process is iterative~\cite{boden2009.pdf} and links ideas between subsequent prompts~\cite{1-s2.0-S0749597817300559-main.pdf}.
Several iterations are often needed to arrive at a subjectively satisfactory result.
The process is also affected by chance and the practitioner's willingness to ``go with the flow'' and allow the conversation with the AI to steer in unanticipated directions.
    For instance, practitioners may adapt their prompts to serendipitously follow opportunities presented by the AI
    (thereby avoiding fixation~\cite{jansson1991.pdf}).
Iteration is also inherently present in how images are created by AI,
as described in the following section.


\newlength{\imagewidthcuration}
\setlength{\imagewidthcuration}{.23\textwidth}

\begin{figure*}[!ht]
\centering
\begin{tabular}{ccccc}
    \includegraphics[width=\imagewidthcuration,height=.21\textwidth]{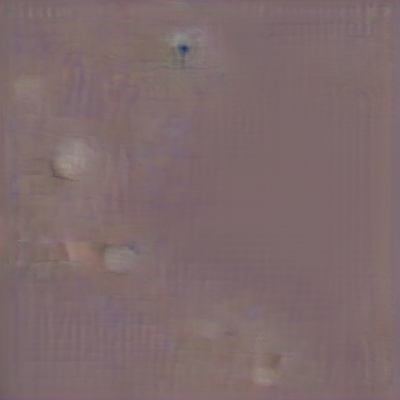}
       &
    \includegraphics[width=\imagewidthcuration,height=.21\textwidth]{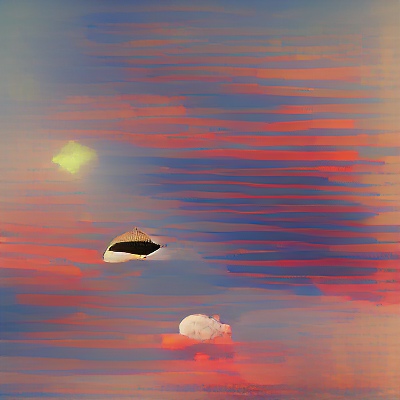}
       &
    \includegraphics[width=\imagewidthcuration,height=.21\textwidth]{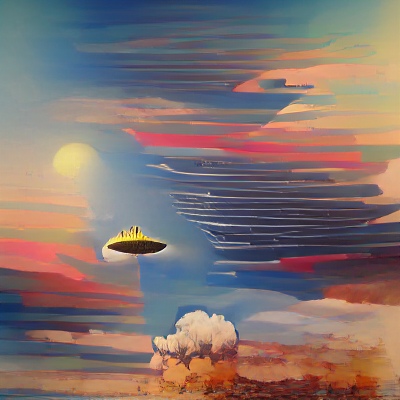}
       &
    \includegraphics[width=\imagewidthcuration,height=.21\textwidth]{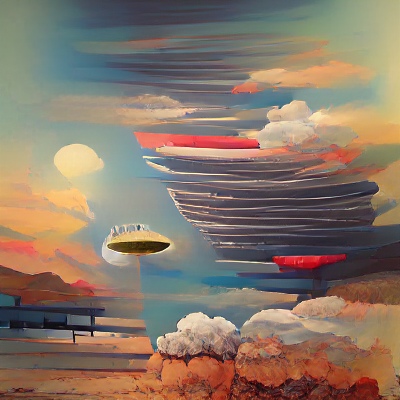}
    \\
        Step 0
       &
        Step 25
       &
        Step 50
       &
        Step 75
    \\
\addlinespace[1ex]

    \includegraphics[width=\imagewidthcuration,height=.21\textwidth]{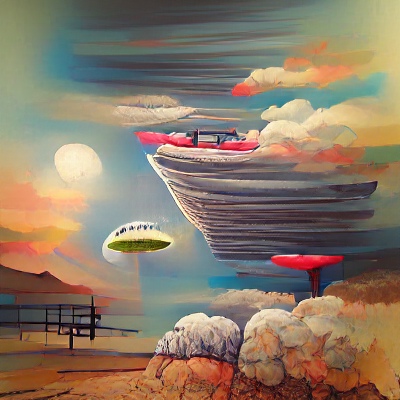}
       &
    \includegraphics[width=\imagewidthcuration,height=.21\textwidth]{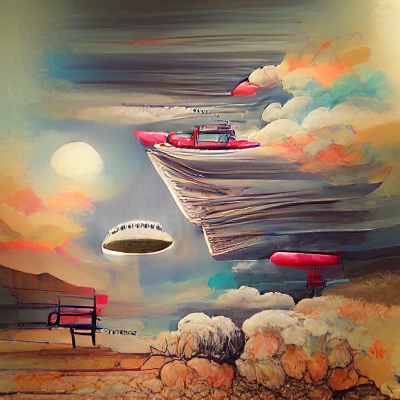}
        &
    \includegraphics[width=\imagewidthcuration,height=.21\textwidth]{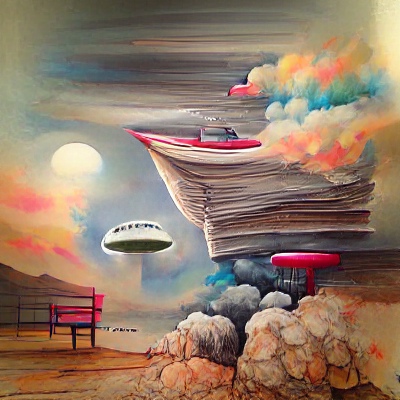}
        &
    \includegraphics[width=\imagewidthcuration,height=.21\textwidth]{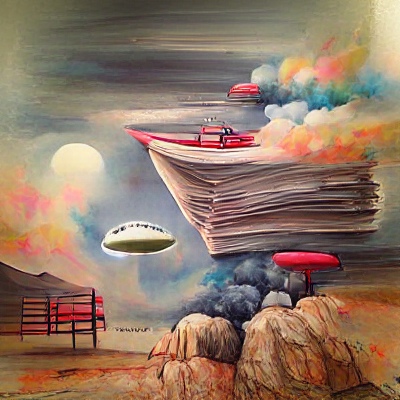}
    \\
        Step 100
       &
        Step 125
        &
        Step 150
        &
        Step 175
\end{tabular}%
  \caption{An example of image-level curation. The image was created with VQGAN-CLIP \cite{VQGANCLIP} with the
  prompt ``ufo landing, fantastic, on Vellum, trending on /r/art.''
  The image slowly emerged from the initial random noise used to seed the GAN (step 0). After 100 steps, unwanted details started to emerge and the author aborted the process after 175 steps.
  The author then selected step 50 as ``best'' (i.e.,  most interesting and artistic) image in the batch.
  }%
  \Description{A demonstration of image-level curation. Several images generated with a text-to-image synthesis system are depicted. It is explained that one of the intermediate steps was chosen as final image.}%
  \label{fig:imagecuration}%
\end{figure*}%
%

\subsubsection{Image-level and portfolio-level curation}%
Images created with text-to-image synthesis do not materialize in their final form. Instead, text-to-image generation systems are iterative and take the previous output as input for the subsequent step. One can therefore visually observe how the images slowly emerge over time.
In this regard, text-to-image synthesis can be compared with the process of developing a Polaroid photograph and the physical processing of film material in a film laboratory (see Figure \ref{fig:imagecuration}).
At any point in this process, the practitioner may decide that the result will not be satisfactory and abort the ``development process.''
    This is the case with GAN-based systems, such as VQGAN--CLIP, which can degenerate to sub-optimal solutions at some point in the image generation process~\cite{VQGANCLIP}. Practitioners therefore often have to select a single image from the set of generated images, and the best image is not necessarily the one generated in the last generation step.
The creativity of text-to-image art, therefore, is also a creativity of \textit{image-level curation} within the wider process of text-to-image generation (see Figure~\ref{fig:imagecuration}).

Further, \textit{portfolio curation} is an important constituent of the creativity of text-to-image art. Not every generated image is up to the standard of the person writing the prompt. Consequently, practitioners of text-to-image art carefully curate a portfolio of their best works, while discarding (or deciding not to publish) other images.
The curative creative aspect of text-to-image art is similar to a photographer's process of weeding out and curating images from a larger collection of images.
Midjourney's community has this curation process built-in. In an online gallery\footnote{
https://www.midjourney.com/app/}, only the images that the Midjourney member chose to upscale are shown by default, thereby reflecting the member's curated works.
Due to curation playing an important role in the image creation process, holistic metrics of curation~\cite{kernecuration1} may be applicable to assess the creativity of the practitioner's portfolio curation.

\subsection{Press: The Emerging Text-to-Image Ecosystem}%
Besides the recent leap in the technical development of text-to-image generation systems, online communities dedicated to text-to-image art are one of the most significant developments. A lively online ecosystem around text-to-image  art has emerged. This ecosystem is enabled by the technologies described in Section~\ref{sec:TTIS} and includes online communities on both general social media platforms (e.g., Twitter and Reddit) and in dedicated online communities. Further, the creative ecosystem includes learning resources, tools, and online services. Together, these online communities and resources contribute to and support the practitioners' creativity~\cite{CSTs,CSTsChung}.



\subsubsection{Communities dedicated to text-to-image art}%
Dedicated online communities have emerged around text-based generative art.
Members in these communities share their prompts and artworks (or even generate images in the community itself). These online communities include, for instance, Reddit's /r/Media\-Synthesis, Twitter, Midjourney~\cite{midjourney}, and many other Discord-based communities. Discord, in particular, has proven useful to the formation of communities as it allows to quickly spin up a chat-based community around a certain topic.
%
The key roles of members in the broader text-to-image online community can be described as follows:%
\begin{enumerate}
    \item The community is enabled by 
    \textit{innovators} who introduce novel techniques and technologies to the community.
    For instance, Ryan Murdock and Katherine Crowson generously shared their software code as open source which has significantly advanced the field of text-to-image generation, as mentioned in Section~\ref{sec:TTIS}.
    \item Not every innovator publishes their code in Colab notebooks. This is where \textit{porters} step in and convert code from repositories on GitHub to executable notebooks on Colab.     Since the field moves quickly, porters often refer to their notebooks as ``quick,'' ``rushed,'' or ``dirty'' implementations to not diminish the effort of the software code's original creator.
    \item \textit{Conservators} seek to preserve the code created by innovators in, for instance, GitHub repositories.
    \item \textit{Service and resource providers} create resources and applications using the open source code provided by innovators.
    \item \textit{Practitioners} (enthusiasts, hobbyists, but also artists) use the notebooks and tools provided by innovators and the systems provided by service providers to generate digital images and artworks.%
\end{enumerate}%
The above roles in the community are fluid. For instance, an innovator may briefly take the role of a porter to make a new technology available in a Colab notebook and advance the field.

Online communities act as a fertile ground for learning the skills necessary to use text-to-image generation systems creatively. One such skill is 
the creative practice of ``prompt engineering'' \cite{gwern,chilton,modifiers}. 
As mentioned in sections~\ref{sec:person} and~\ref{sec:process}, prompt engineering is a skill learned with experimentation and by engaging with prior work~\cite{p380-kerne.pdf}.
Because enthusiastic creators share their prompts and practices, online communities are a fertile ground for novice practitioners to learn from other members of the community in order to overcome the learning curve associated with prompt engineering. But while online communities act as catalysts for creativity, the distributed nature and sequential presentation of messages in the communities make it difficult to pursue specific learning goals (such as learning about prompt modifiers). Consequently, practitioners and online communities increasingly strive to create dedicated resources that better support their members, as described in the following section.

\subsubsection{Dedicated learning resources}%
The Midjourney community is on its way to becoming a cultural hub due to its attention to features that benefit the community as a whole. Midjourney has, for instance, introduced member profiles, exploratory maps, and information on style modifiers that help its members to explore and learn the craft of prompt engineering.
Besides online communities, community members have created learning resources on prompt engineering.
    For instance, practitioners have created a number of resources for specific goals, such as style experimentation (e.g., \cite{DiscoDiffusionArtiststudies,artstudies}) and teaching prompt engineering (e.g., the DALL-E prompt book~\cite{dallepromptbook}).
Some of these resources follow an approach similar to the systematic experimentation by \citeauthor{chilton}~\cite{chilton}.
    For instance, practitioners have shared their experiments on artist names that may be used as style modifiers in input prompts.
    Such resources include, for instance, Remi Durant's artist studies (a list of artist names for generating images in a certain artistic style) \cite{artstudies} and
    Harmeet Gabha's list of Disco Diffusion modifiers \cite{disco-diffusion-modifiers}.
    Hub pages are another type of resource. Since not all Colab notebooks are indexed in search engines, hub pages were created in an effort to make the quickly growing field of text-to-image generation discoverable. Hub pages act as indexes that provide collections of links to Colab notebooks.
    Examples of hub page are
        Lj Miranda's list of {VQGAN--CLIP} implementations~\cite{vqganlist},
        pharmapsychotic's ``Tools and Resources for AI Art''~\cite{pharmapsychotic}, and
        Reddit user Wiskkey's ``List of Stable Diffusion systems'' \cite{wiskkey}.%
%
\subsubsection{Tools and services}%
A growing number of services and tools are emerging on the Web
to support practitioners in creating text-to-image art.
Besides the aforementioned Midjourney community, which has amassed a membership approaching 1 million community members~\cite{midjourney-growth},
such services include
    Artbreeder\footnote{https://www.artbreeder.com} (a service that allows its users to fork other users' AI-generated artworks),
    NightCafe\footnote{https://creator.nightcafe.studio}, 
    Pollinations\footnote{https://pollinations.ai},
    Visions of Chaos\footnote{https://softology.pro/voc.htm} (a desktop software for experimenting with different deep learning models),
    WOMBO Dream\footnote{https://www.wombo.art} and
    Starry AI\footnote{https://www.starryai.com} (mobile applications for generating text-to-image art),
    and multimodal.art\footnote{https://multimodal.art} (a user interface built on top of Google's Colab).
Tools include, for instance, CLIP retrieval\footnote{https://rom1504.github.io/clip-retrieval} that helps to find out how the CLIP model understands a text prompt, and CLIP prefix captioning\footnote{https://github.com/rmokady/CLIP\_prefix\_caption}, a model that applies CLIP for captioning images with the aim of helping to produce better prompts from this initial image.
The growing ecosystem of services and tools assists in making text-to-image generation more accessible to non-technically mind\-ed practitioners. The ecosystem also acts as a catalyst that draws in new practitioners and advances the field as a whole.
It is the specific combination of people, technology, services, tools, and resources that formed a healthy ecosystem for the text-to-image art community to thrive.

\section{Discussion}%
\label{sec:discussion}%
Art is subjective and, first and foremost, meaningful to its creator (according to the small-c paradigm of creativity). We humans enjoy pretty things and intrinsic motivation is one key driver to engage in the production of text-to-image art.
However, some practitioners also pursue other interests -- the money, fame, and glory 
of successful artists (Big-C perspective).
Artworks synthesized with text-to-image generation systems (or edits thereof) are being sold as NFT (non-fungible cryptographic tokens built on blockchain-based technology) in online marketplaces dedicated to digital art~\cite{NFTArt}.
In combination with NFT technology, text-to-image art has, therefore, contributed to boosting the digital creative economy.

With text-to-image generation systems, anybody can create art.
The easy access to this technology makes it possible for anybody to sell artworks, for instance, as NFTs.
This raises questions about the level and nature of the human creativity involved in text-to-image synthesis.
Melvin Rhodes wrote in 1961 that each ``strand'' of the four-part model of creativity ``has unique identity academically, but only in unity do the four strands operate functionally''~\cite{4P}.
By focusing on the product-centered op\-era\-tion\-al\-iz\-ation of the concept of creativity, 
science seems to have lost track of this big picture. We obsess so much over p-values that we seem to have forgotten that human creativity can, as demonstrated in this paper, be deeply rooted in factors that are not covered by the ``standard'' product-based definition of creativity \cite{2012RuncoJaegerStandardDefinition.pdf}.
But not only is science's one-sided view a subject for concern, but there are also concrete challenges when assessing the creativity of text-to-image art.


\subsection{Challenges for Evaluating the Creativity of Text-to-Image Art}%
Text-to-image art is the result of an opaque process. When evaluating the creativity of text-to-image art, we run into three challenges 
    (system, prompts, and process)
related to information asymmetries between the creator and the viewer of the artwork.

\subsubsection{System}
Based on the digital image alone, we can infer little about the generative system used to synthesize the image.
Some text-to-image generation systems have dozens of configuration parameters that can individually be adjusted to optimize the results.
Clearly, knowing and adjusting the configuration parameters plays a crucial role in distinguishing skillful and purposeful mastery of text-to-image art from beginner's use with default settings.

\subsubsection{Prompts}
Many practitioners of text-based generative art do not share their prompts. Among these practitioners, prompts may be considered as a trade secret~-- especially among practitioners who pursue commercial interests. Therefore, when we encounter an artwork posted on social media, we often do not know the prompt used to generate the image. In other cases, the creator may have shared the prompt (or parts thereof), but not the full set of prompt modifiers.
Prompt modifiers are keywords that determine the style and quality of the generated image \cite{modifiers}. Knowing the full set of prompt modifiers is therefore crucial for understanding how the image was generated.

Further complicating the assessment of creativity is the fact that some systems can accept images as input prompts in addition~-- or as substitute~-- to textual prompts.
    Text-to-image generation systems are inherently iterative by design. 
    Images can therefore be used as initial input for the system to seed the first iteration.
        Such initial images can, for instance, be used to direct the image's scene composition or to manipulate and distort existing images~\cite{VQGANCLIP}.
    Further, some text-to-image generation systems accept one or several target images as prompts. These visual prompts act like text prompts in that they provide a (visual) guide and a target for the system to optimize its losses.
It is difficult and often impossible to tell what kind of inputs were used (i.e., text, images, or combinations thereof) to generate an image.

\subsubsection{Process}
From the digital image alone, we know little about the process used to create the image.
For instance, we do not know how the practitioner came up with the prompt.
A text-to-image artwork could, as demonstrated in Section~\ref{sec:edgecase}, be the result of a near skill-free interaction with the text-to-image generation system, the practitioner could have gotten lucky on the first try, or the practitioner could have just copied someone else's prompt or prompt modifiers.
    On the other hand, the artwork can also be the result of an arduous iterative process~-- a focused conversation with the AI system~-- that constitutes the process of prompt engineering for text-to-image generation.

For assessing the full extent of human creativity involved in text-to-image art, we need information on all three of the above aspects.

\subsection{Opportunities for Future Research on Text-to-Image Generation}%
From the perspective of Human-Centered Computing, the phenomena described in this paper are a ``blend of technology, humans, and community''~\cite{p32-guzdial.pdf}.
This section describes future research opportunities structured around these three aspects.

\subsubsection{Technology -- better understanding the user's intent}%
Practitioners interact with text-to-image generation systems via textual prompts.
Today, the systems' understanding of textual input prompts is far from perfect and the concepts learned by the neural network may not correspond to concepts in our visual world~\cite{aliendream}.
Natural language understanding~(NLU) is a part of the field of natural language processing~(NLP) concerned with making machines understand human language. In the context of text-to-image generation, advances in natural language understanding are needed to minimize the gap between the user's implicit intent, the explicitly stated textual prompt, and the visual output of the
text-to-image generation system.

\subsubsection{Community -- understanding the practice of prompt engineering and the design of co-creative systems}%
Text-guided generative art is among the first areas in which the practice of prompt engineering is being exercised ``in the wild.''
Since text-to-image generation involves natural language, practices and lessons learned from
large-scale language models (LLMs) -- and in particular {GPT-3} \cite{2005.14165.pdf,2102.07350.pdf} -- could be applied to prompt engineering for text-to-image synthesis.
    For instance, {Open\-AI} maintains a collection of input templates for use with {GPT-3}~\cite{gpt-3-templates} and similar resources and tools are emerging in the text-to-image art community.
Some first design guidelines for prompt engineering in the context of text-to-image generation have been published~\cite{chilton}, but there is still much to learn about the creative practices emerging in the communities around text-to-image art and how systems can be designed to support practitioners.
%
%
%
For instance, the interaction with text-to-image generative systems takes place through web-based interfaces. Research on these user interfaces could inform the design of novel co-creative systems~\cite{3519026.pdf} and creativity support tools~\cite{CSTs,CSTsChung,dc2s2,crowdpoweredCST}.
    For instance, dedicated user interfaces have been developed aiming at better supporting practitioners in creating text-to-image art (e.g., the MindsEye interface provided by multimodal.art\footnote{https://multimodal.art}).


\subsubsection{Humans -- co-creation and its effect on society}%
We are witness to an emergence of a new type of creative economy that requires no prior experience or skill~-- a democratization of art and creative production. Some say that this paradigm shift is of a magnitude even greater than the shift when photography was first introduced~\cite{arts-07-00018-v3.pdf}.
Back then, artists saw their livelihood being threatened by photography and~-- in fear of photography superseding the art of painting~-- ridiculed photography as not being art \cite{photographynotart}.
In hindsight, photography unlocked a vast amount of creativity that was traditionally corralled and coveted by painters. While some artists were negatively affected (especially those who specialized in producing photo-realistic portraits), the invention of photo\-graphy opened opportunities to many more.
It is likely that generative AI will lead to a similar singularity event in our human history.
This change could deeply affect the role of the creator and our relation to images and creative work could change as a result, with implications for society as a whole.

AI-based systems, as an extension of our human cognitive workbench, have the potential to augment and support our creativity \cite{CSTs,engelbart1962}.
But our increasing interaction with and involvement in co-creative systems may also have an effect on our behavior, language, knowledge, and skills.
Similar to how the World Wide Web changes society (as much as society changes the Web), interaction with AI in natural language will shape the future digital society, the way we interact with computers, and the way we work.
We may, in the future, even see a bidirectional effect of prompt engineering on society: As we become more accustomed to phrasing our inputs for foundation-scale machine learning models \cite{foundationmodels}, the way we use language and communicate with other humans may consequently change as well.

\section{Conclusion}%
\label{sec:conclusion}%
The product-centered view of creativity is not enough to fully assess the creativity of art synthesized with text-to-image generation systems.
This work demonstrated this point with a scenario in which the product-centered definition of creativity fails to capture the full extent of the human creativity involved in text-to-image generation.
The paper argued that in order to evaluate the full extent of human creativity involved in text-to-image generation, we need to look not only at the digital images as outcome, but also the creator's process and environment. Drawing on Rhodes' conceptual ``four P'' model, this paper expanded on the nature of human creativity involved in text-to-image generation.
    Image-level and portfolio-level curation, for instance, are ways in which creators express their creativity.
The human creativity of text-to-image art lies in the interaction with text-to-image generation systems and in online communities.
Increasingly, online communities act as a catalyst for learning the craft of ``prompt engineering.'' This craft, practiced by members of a growing online community of text-to-image practitioners, offers an exciting area of research in HCI.






\balance
\bibliographystyle{ACM-Reference-Format}
\bibliography{paper}


\end{document}